\documentclass[%
    aip,
    superscriptaddress,
    reprint,
    graphicx,
    nofootinbib,
    aps,
    amsmath,amssymb,
    floatfix,
]{revtex4-2}

\draft 

\usepackage{graphicx}
\usepackage{dcolumn}
\usepackage{bm}
\usepackage[utf8]{inputenc}
\usepackage[T1]{fontenc}
\usepackage{etoolbox}
\usepackage{makecell}
\usepackage{multirow}
\usepackage{threeparttable}
\usepackage[symbol]{footmisc}

\begin{document}


\title{Probing autoionization decay lifetimes of the $\mathbf{4d^{-1}6\boldsymbol{\ell}}$ core-excited states in xenon using attosecond noncollinear four-wave-mixing spectroscopy} 



\author{Nicolette G. Puskar}
    \altaffiliation[nicolette.puskar@berkeley.edu; ]{These authors contributed equally to this work.}
\affiliation{Department of Chemistry, University of California, Berkeley, California 94720, USA}
\affiliation{Chemical Sciences Division, Lawrence Berkeley National Laboratory, Berkeley, California 94720, USA}

\author{Patrick Rupprecht}
    \altaffiliation[prupprecht@lbl.gov; ]{These authors contributed equally to this work.}
\affiliation{Department of Chemistry, University of California, Berkeley, California 94720, USA}
\affiliation{Chemical Sciences Division, Lawrence Berkeley National Laboratory, Berkeley, California 94720, USA}

\author{Jan Dvo\v{r}\'{a}k}
\affiliation{Chemical Sciences Division, Lawrence Berkeley National Laboratory, Berkeley, California 94720, USA}

\author{Yen-Cheng Lin}
    \altaffiliation[Present address: ]{Sandia National Laboratories, Livermore, CA 94550, USA}
\affiliation{Department of Chemistry, University of California, Berkeley, California 94720, USA}
\affiliation{Chemical Sciences Division, Lawrence Berkeley National Laboratory, Berkeley, California 94720, USA}

\author{Avery E. Greene}
    \altaffiliation[Present address: ]{Department of Chemistry, University of Wisconsin-Madison, Madison, Wisconsin 53706, USA}
\affiliation{Department of Chemistry, University of California, Berkeley, California 94720, USA}

\author{Robert R. Lucchese}
\affiliation{Chemical Sciences Division, Lawrence Berkeley National Laboratory, Berkeley, California 94720, USA}

\author{C. William McCurdy}
\affiliation{Chemical Sciences Division, Lawrence Berkeley National Laboratory, Berkeley, California 94720, USA}
\affiliation{Department of Chemistry, University of California, Davis, California 95616, USA}

\author{Stephen R. Leone}
\affiliation{Department of Chemistry, University of California, Berkeley, California 94720, USA}
\affiliation{Chemical Sciences Division, Lawrence Berkeley National Laboratory, Berkeley, California 94720, USA}
\affiliation{Department of Physics, University of California, Berkeley, California 94720, USA}

\author{Daniel M. Neumark}
\affiliation{Department of Chemistry, University of California, Berkeley, California 94720, USA}
\affiliation{Chemical Sciences Division, Lawrence Berkeley National Laboratory, Berkeley, California 94720, USA}
    \email{dneumark@berkeley.edu}

\date{\today}

\begin{abstract}
The decay of core-excited states is a sensitive probe of autoionization dynamics and correlation effects in many-electron systems, occurring on the fastest timescales. Xenon, with its dense manifold of autoionizing resonances that can be coupled with near-infrared light, provides a platform to investigate these processes. In this work, the autoionization decay lifetimes of $4d^{-1}6\ell$ $(\ell = s, p, d,  ...)$ core-excited states in xenon atoms are probed with extreme ultraviolet (XUV) attosecond noncollinear four-wave-mixing (FWM) spectroscopy. The $4d^{-1}_{\{5/2,\, 3/2\}}6p$ XUV-bright states (optically dipole allowed) exhibit decay lifetimes of $\sim$6 fs, which is consistent with spectator-type decay. In contrast, the $4d^{-1}_{\{5/2,\, 3/2\}}6s$ and $4d^{-1}_{\{5/2,\, 3/2\}}6d$ XUV-dark states (optically dipole forbidden) show longer decay lifetimes of $\sim$20~fs. Photoionization calculations confirm that all core-hole states with $4d$ character should decay via spectator channels in $\leq$ 6~fs, suggesting that the apparent longer dark state decay times arise from an alternative mechanism. A few-level simulation of the FWM process shows that the inclusion of a nearby, longer-lived dark state can mimic the experimental FWM signal, suggesting population cycling with a second electronic state with non-$4d$ character. \textit{Ab-initio} calculations support the presence of such multi-electron excited states in the 60–70~eV range. These results demonstrate that FWM signals can encode coupled-state dynamics when probing complex systems, highlighting the importance of combining theoretical and experimental approaches to disentangle accurate core-level decay pathways and lifetimes.
\end{abstract}


\maketitle 

\section{Introduction}
\label{sec:introduction}

\begin{figure*}
\includegraphics{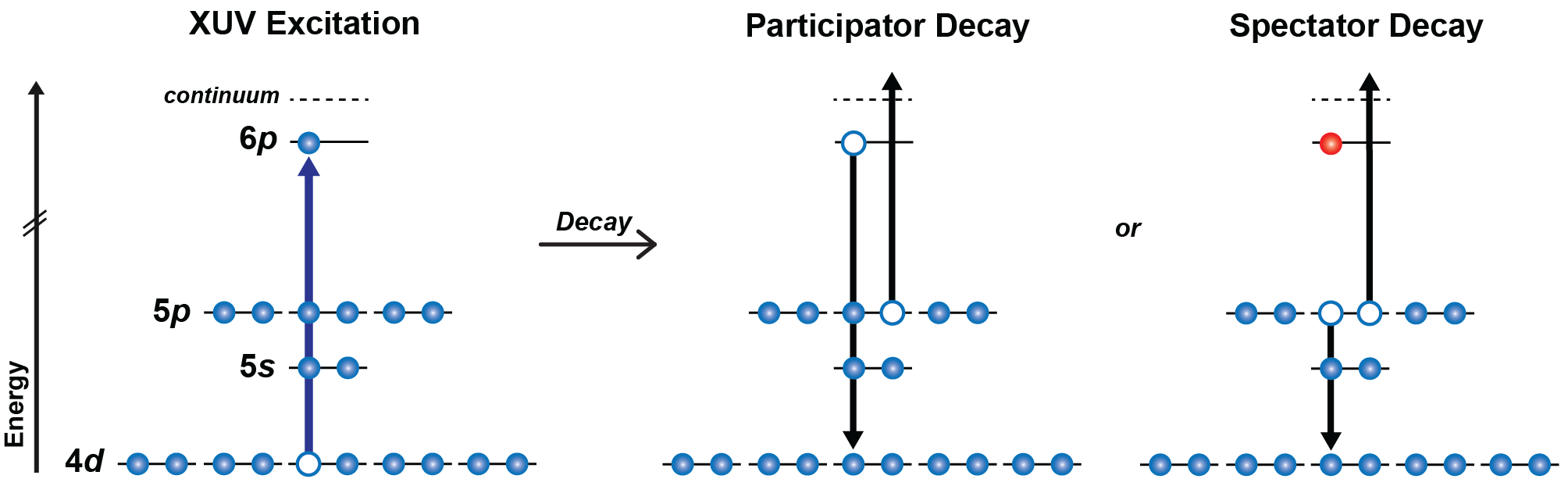}
\caption{Examples of participator- and spectator-type decay mechanisms after a $4d \rightarrow 6p$ XUV excitation (electrons are blue spheres, holes are white circles). During participator decay, the electron that is initially excited into the $6p$ orbital will "participate" in the autoionizing process by either filling the $4d$ hole or by being ionized. In this figure, the $6p$ electron is shown to fill the $4d$ hole while a $5p$ electron ionizes. If the $6p$ electron were to ionize instead, the $4d$ hole would then be filled by either a $5s$ or a $5p$ electron. During spectator decay, the initially excited electron remains in the $6p$ orbital (red sphere) while other electrons reconfigure to fill the $4d$ hole and ionize. In this figure, the dominant spectator transition (and overall dominant decay channel) $4d^{-1}np \rightarrow 5p^{-2}np$ as found by Aksela \textit{et al.}\cite{aksela_electron_1994} is depicted.
} 
\label{fig:ao_diagram}
\end{figure*}

Autoionization is a process by which highly excited electronic states of atoms, molecules, and solids can decay after photo-excitation with extreme ultraviolet or x-ray light.\cite{russek_auto-ionization_1968,rehmus_mechanism_1981,bader_autoionization_1986} In atoms, after a photon excites an electron from a core level to some discrete state above the ionization threshold, the electronic structure rearranges upon relaxation and one electron fills the original vacancy while another is spontaneously ejected.\cite{shenstone_ultra-ionization_1931,beutler_uber_1935} This Auger-Meitner decay is driven by electron correlation. In the case of a "participator" transition, the initially excited electron is involved in the decay by either filling the hole or being ionized. Conversely, a "spectator" transition occurs when the initially excited electron remains in the state to which it was excited while other electrons rearrange, leaving the cation in an excited electronic state (Fig.~\ref{fig:ao_diagram}).\cite{armen_valence_1991} In core-level excitations, as opposed to valence or inner-valence-level transitions, the excited electron is typically more likely to act as a spectator.\cite{fano_line_1965} One such excitation that has served as a model for understanding the decay dynamics of core-excited states is the promotion of a $4d$ electron in xenon into its $np$ Rydberg series. In this article, the lifetimes of $4d$ core-excited states in Xe are directly measured with attosecond noncollinear four-wave mixing spectroscopy. 

The $4d$ core level is spin-orbit split, resulting in two identifiable series: the $4d_{5/2}^{9}5s^{2}5p^{6}[np]$ and $4d_{3/2}^{9}5s^{2}5p^{6}[np]$ Rydberg series\cite{kramida_nist_2024}, which have been investigated in a series of frequency-domain experiments. The energy positions and natural linewidths of the $4d_{\{5/2, \, 3/2\}}^{-1}np$ states were first reported by Ederer and Manalis\cite{ederer_photoabsorption_1975} using synchrotron radiation, and were followed shortly after by King \textit{et al.}\cite{king_investigation_1977} using high-energy electron energy-loss spectroscopy (EELS). These experiments have since been repeated with synchrotron radiation by Masui \textit{et al.}\cite{masui_new_1995} and Sairanen \textit{et al.}\cite{sairanen_high-resolution_1996}, in addition to a recent EELS study performed by Zhang \textit{et al.}\cite{zhang_fast-electron-impact_2015} Of interest to this study are the results for the natural linewidths of these states because not only do they provide insight as to whether the state undergoes participator or spectator decay, they also recover the natural decay lifetime of the state through the lifetime and energy width relationship. 

The studies mentioned above have reported linewidths with a range of 90 -- 128 meV and 104 -- 133 meV for the $4d_{5/2}^{-1}np$ and $4d_{3/2}^{-1}np$ states for $n=\{6,7,8\}$, respectively, which convert to natural decay lifetimes of 5 -- 7 fs (see Supplementary Material (SM) Table~I). The near-constant linewidths as $n$ changes in the Rydberg series suggest that the Xe $4d$ electrons undergo spectator-type decay. This idea was demonstrated by Fano and Cooper\cite{fano_line_1965}, who showed that participator transitions lead to linewidth variation on the order of approximately $n^3$ and that the linewidths of spectator transitions are independent of $n$. Aksela \textit{et al.}\cite{aksela_correlation_1995} conducted a set of experiments using Auger electron spectroscopy, which not only verified the dominance of the spectator transition, but also asserted that $4d^{-1}np \rightarrow 5p^{-2}np$ was the dominant decay channel.\cite{aksela_electron_1994} In this primary channel, after the $4d$ electron is excited into any $np$ Rydberg orbital, the Rydberg orbital remains filled while two $5p$ electrons rearrange during decay, with one filling the $4d$ core-hole and one undergoing ionization (Fig.~\ref{fig:ao_diagram}). 

\begin{figure}
\includegraphics{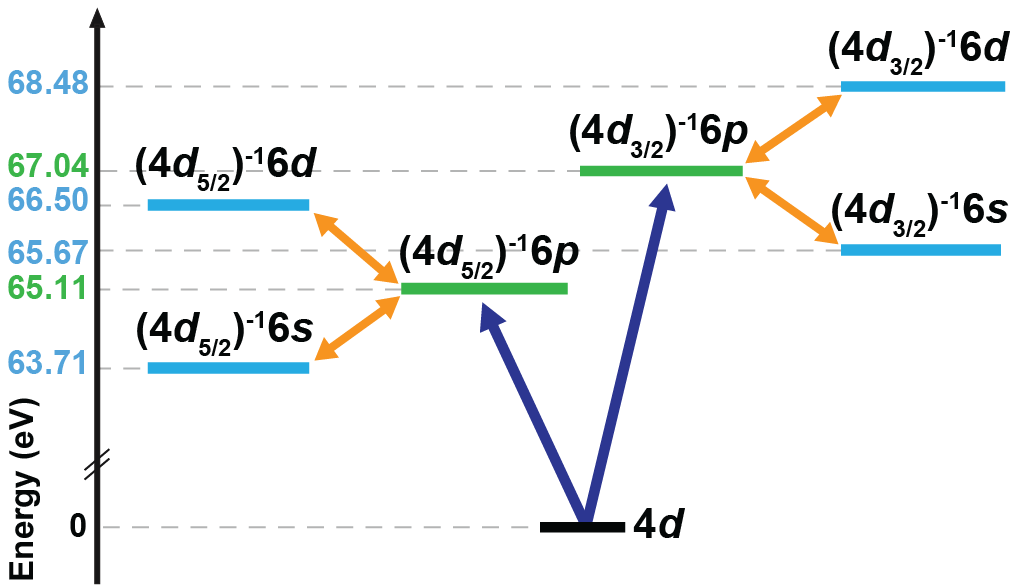}
\caption{Energy level diagram of xenon $4d^{-1}6\ell$ states. A ground state $4d$ electron (black line) is excited with XUV light (purple arrows) into the spin-orbit split $4d_{5/2}^{-1}6p$ and $4d_{3/2}^{-1}6p$ XUV-bright states (green lines). The left side of the diagram depicts the nearby $4d_{5/2}^{-1}6s$ and $4d_{5/2}^{-1}6d$ XUV-dark states (blue lines) that can be reached with the NIR pulses (orange arrows), while the right side depicts the same for $4d_{3/2}^{-1}6p$. The energies of each state are given on the \textit{y}-axis in eV units, and were taken from Kolbasova \textit{et al}.\cite{kolbasova_probing_2021}} 
\label{fig:xenon_energy_level_diagram}
\end{figure}

While the decay mechanisms for the optically accessible $4d^{-1}np$ bright states are reasonably well-understood, open questions remain about the decay of the $4d^{-1}n\{s,\, d\}$ dark states, which are optically inaccessible from the ground state (Fig.~\ref{fig:xenon_energy_level_diagram}). While the EELS study by Zhang \textit{et al.}\cite{zhang_fast-electron-impact_2015} yielded the energy positions for the $4d_{\{5/2,3/2\}}^{-1}6d$ and $4d^{-1}_{5/2}6s$ dark states, only the natural linewidth of the $4d^{-1}_{5/2}6s$ state was reported at 102~meV, which converts to a 6.5 fs natural decay lifetime. To the best of the authors' knowledge, no other experimental or theoretical investigations have been published about the decay dynamics of these specific dark states so far. 

One method capable of accessing dark-state dynamics is attosecond noncollinear four-wave-mixing (FWM) spectroscopy, an explicitly time-domain experiment. FWM is an all-optical technique that can time-resolve electronic dynamics of both bright and dark states. It can probe few-femtosecond decay dynamics in atoms\cite{rupprecht_extracting_2024, puskar_measuring_2023, fidler_autoionization_2019, marroux_multidimensional_2018}, molecules\cite{fidler_state-selective_2022, lin_coupled_2021}, and solids\cite{gaynor_solid_2021} in a quantum-state-specific manner. It achieves this specificity through utilization of a noncollinear beam geometry between an extreme-ultraviolet (XUV) pulse and two near-infrared (NIR) pulses, which results in background-free, spatially isolated FWM signals. In this study, a 40 -- 70 eV XUV pulse excites a $4d$ electron into the $4d^{-1}6p$ states (Fig.~\ref{fig:xenon_energy_level_diagram}). The two NIR pulses can then couple the $4d^{-1}6p$ XUV-bright states to the $4d^{-1}6s$ and $4d^{-1}6d$ XUV-dark states in a V- or $\Lambda$-type coupling scheme.\cite{puskar_measuring_2023} With this technique, the core-excited bright- and dark-state dynamics of xenon can be probed in the time-domain for the first time.

In this work, attosecond FWM is used to characterize sub-6~fs lifetimes of core-level excited states in xenon. The autoionizing decay lifetimes of the spin-orbit-split $4d^{-1}_{\{5/2, \, 3/2\}}6p$ bright states are measured directly in the time domain and are found to be 4.8 $\pm$ 1.0 fs for $4d^{-1}_{5/2}6p$ and 5.8 $\pm$ 1.3 fs for $4d^{-1}_{3/2}6p$, in agreement with spectrally-deduced values from synchrotron measurements as well as with our calculations (Sec. \ref{subsec:measured decay lifetimes}). 
However, FWM measurements targeting the dipole-inaccessible $4d^{-1}_{\{5/2, \, 3/2\}}6s$ and $4d^{-1}_{\{5/2, \, 3/2\}}6d$ dark states display puzzlingly longer decay lifetimes of $\sim20$~fs. 
Our \textit{ab-initio} photoionization calculations suggest that the lifetimes of all $4d^{-1}n\ell$ core-hole states decay in $\leq 6$~fs (Sec. \ref{sunbsec:photoion_decay}).
Simulating the FWM experiments with a few-level time-dependent Schrödinger equation (TDSE) calculation indicates that another non-$4d$-hole character dark-state with a lifetime of about 20~fs can contribute to the FWM dark-state dynamics, while such states do not affect the results for the bright-state lifetimes (Sec. \ref{sec:FWM_simulation:_results}).
Indeed, \textit{ab-initio} restricted active space configuration interaction (RASCI) calculations reveal the presence of many multi-electronic excitation states in the 60 to 70~eV energy region that do not have a $4d$ hole (Sec. \ref{subsec:darkState_decay}).

\section{Experimental Methods}
\label{sec:experimentalmethods}

\begin{figure}
\includegraphics{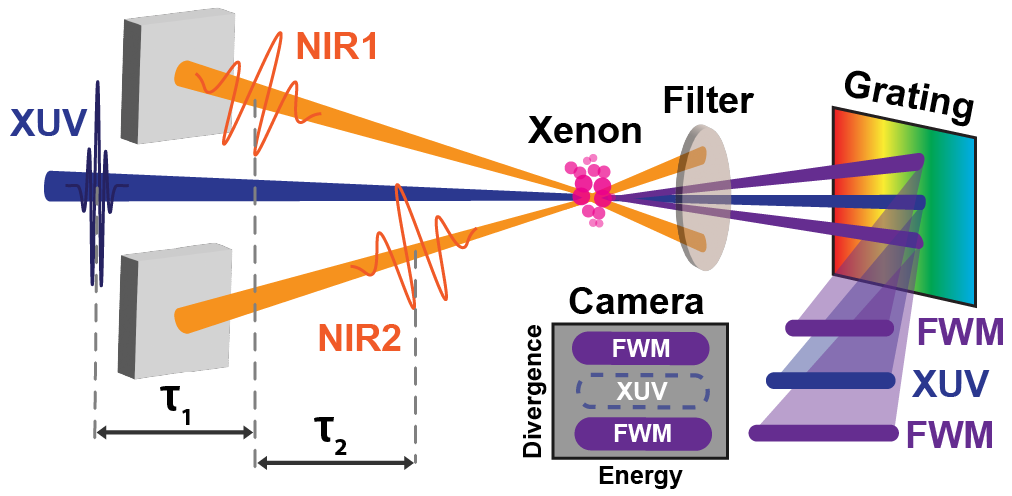}
\caption{Experimental schematic of attosecond noncollinear FWM spectroscopy. One attosecond XUV pulse interacts with two time-delayable NIR pulses at the xenon sample in a third-order nonlinear process and generates FWM signals, which are spatially isolated due to the noncollinear geometry. An Al filter attenuates the NIR signals while the FWM and transmitted XUV beams propagate to the grating, which spectrally disperses the signal onto the CCD camera. The FWM signals are spatially distinct on the divergence axis, which can be varied according to Eq.~\eqref{eqn:divergence_angle}. 
}
\label{fig:experimental_fwm_geometry}
\end{figure}

The apparatus for attosecond noncollinear FWM has been described previously\cite{puskar_measuring_2023, rupprecht_extracting_2024} and is summarized in Fig.~\ref{fig:experimental_fwm_geometry}. A commercial Titanium:Sapphire laser system (Legend Elite Duo HE, Coherent) produces 40 fs pulses centered at 800 nm at a 1 kHz repetition rate. The pulses are spectrally broadened by focusing 4 mJ of the output power into a 3-meter-long stretched hollow-core fiber system (few-cycle, Inc.) pressurized with neon gas, and then temporally compressed with chirped mirrors (PC70, Ultrafast Innovations), resulting in 8~fs NIR pulses (see SM Fig.~1). A 70:30 beamsplitter is then used to divide the beam, where 70\% is used to generate attosecond XUV pulses and 30\% creates the two NIR wave-mixing beams. 

The short attosecond pulse trains are created via high-harmonic generation in argon gas, which results in a broadband XUV spectrum ranging from 40 -- 70~eV. Afterwards the copropagating NIR light is removed with a 0.2~µm aluminum (Al) filter (Lebow). The XUV pulses are then focused into the sample cell with a 10$^{\circ}$ grazing-angle gold-coated toroidal mirror (ARW Optical Corp.). The NIR wave-mixing beams are created by first introducing one piezoelectric delay stage into the beam, followed by a 50:50 beamsplitter, and then adding a second piezoelectric delay stage into only one of the two newly created NIR beams. This technique generates two NIR beams that can be independently delayed with respect to the XUV beam. Finally, the two NIR wave-mixing beams are focused into the sample cell in a noncollinear geometry as shown in Fig.~\ref{fig:experimental_fwm_geometry}. 

The spatial and temporal overlap of the XUV and two NIR pulses is determined with an in-vacuum beta barium borate crystal by optimizing second harmonic generation signals generated by the noncollinear beams. After the FWM signal is generated via the $\chi^{(3)}$ nonlinear process, the XUV and FWM signals are frequency dispersed by a gold-coated variable line space grating (Hitachi) while the two NIR beams are attenuated by a second 0.2~µm Al filter. The resulting spectra are then recorded with an XUV CCD camera (Pixis XO 400B, Princeton Instruments) where the photon energy is read along the \textit{x}-axis and the phase-matching divergence angle is read along the \textit{y}-axis. 

The noncollinear geometry of the FWM allows the FWM signal to be background free. In a V- or $\Lambda$-type coupling scheme, an XUV photon first excites an electron to an XUV-bright state. An NIR1 (NIR2) photon will then resonantly dipole-couple the electron to a nearby XUV-dark state, after which an NIR2 (NIR1) photon couples the electron back down to the original XUV-bright state (Fig.~\ref{fig:xenon_energy_level_diagram}). This results in the emission of FWM signal according to the phase-matching condition: 

\begin{equation}
      \vec{k}_\mathrm{FWM} = \vec{k}_\mathrm{XUV} \pm \vec{k}_\mathrm{NIR1} \mp \vec{k}_\mathrm{NIR2}\:,
\label{eqn:phase_matching}
\end{equation}

\noindent where the wave vector of the FWM signal ($\vec{k}_\mathrm{FWM}$) is governed by the sum and difference of the wave vectors of the XUV ($\vec{k}_\mathrm{XUV}$), NIR1 ($\vec{k}_\mathrm{NIR1}$), and NIR2 ($\vec{k}_\mathrm{NIR2}$) beams.

The divergence angle of the emitted FWM signal can be manipulated to maximize signal isolation away from the XUV spectrum and is given by the following relationship: 

\begin{equation}
    \phi_{\mathrm{div}, V, \Lambda} \approx \pm \frac{E_\mathrm{NIR1}\theta_\mathrm{NIR1} + E_\mathrm{NIR2}\theta_\mathrm{NIR2}}{E_\mathrm{XUV}}\:,
\label{eqn:divergence_angle}
\end{equation}

\noindent where $E_\mathrm{NIR1}$ and $E_\mathrm{NIR2}$ are the photon energies of the NIR1 and NIR2 pulses, $E_\mathrm{XUV}$ is the photon energy of the XUV pulse, and $\theta_\mathrm{NIR1}$ and $\theta_\mathrm{NIR2}$ are the angles of the two incident NIR pulses with respect to the XUV pulse. In the experiments described here, $\theta_\mathrm{NIR1} = \theta_\mathrm{NIR2}$ = \mbox{26 mrad} ($1.5^{\circ}$), $E_\mathrm{NIR1} = E_\mathrm{NIR2}$ = 1.55 eV (800 nm), and $E_\mathrm{XUV} \cong$ 65 eV, resulting in a FWM divergence angle of approximately 1.24 mrad ($0.07^{\circ}$).

Because noncollinear FWM spectroscopy utilizes three pulses, there are two time delays that can be controlled during experiments: the first time delay ($\tau_{1}$) is between the XUV and NIR1 pulses, while the second ($\tau_{2}$) is between the NIR1 and NIR2 pulses. Selecting the appropriate order of the pulses, or pulse sequence, reveals the dynamic information of either XUV-bright states or XUV-dark states.\cite{puskar_measuring_2023} To obtain information from XUV-bright states, a "bright-state scan" is employed, in which the NIR1 and NIR2 pulses are temporally overlapped and delayed simultaneously with respect to the XUV pulse \mbox{($\tau_{2}$ = 0)}. To obtain information from XUV-dark states, a "dark-state scan" is used, where the NIR1 pulse is temporally overlapped with the XUV pulse \mbox{($\tau_{1}$ = 0)} and the NIR2 pulse is delayed independently. In other words, bright-state scans yield bright-state dynamics, and likewise, dark-state scans yield dark-state dynamics.

The phase matching condition given in Eq.~\eqref{eqn:divergence_angle} explains why these two pulse sequences probe different state dynamics within the emitted FWM signal. Due to the presence of XUV-dark states at higher ($4d^{-1}6d$) and lower energies ($4d^{-1}6s$) compared to the targeted XUV-bright state ($4d^{-1}6p$), FWM signals appear at both positive (upper FWM) and negative (lower FWM) divergence angles in time overlap ($\tau_{1} = \tau_{2} = 0$). Since the NIR1 and NIR2 pulses are delayed simultaneously in a bright-state scan, the upper and lower FWM signals contain equivalent information about the XUV-bright $6p$ state. However, only the NIR2 pulse is delayed in a dark-state scan, and therefore the information symmetry is broken. As a result, the upper and lower FWM signals contain information about different states. In the case of xenon, the $4d^{-1}6s$ state dynamics would be revealed in the upper FWM signal while the $4d^{-1}6d$ state dynamics would be revealed in the lower FWM signal.

\section{Theoretical Methods}

To gain deeper insights into the underlying physics of this experiment and the decay dynamics of core-level excited xenon, theoretical efforts on both an \textit{ab-initio} level and an effective few-electronic states level have been undertaken. The following \textit{ab-initio} photoionization calculations aim to extract the decay lifetimes from photoionization cross sections, while a few-level TDSE calculation simulates the FWM signal itself.

\label{sec:theoreticalmethods}

\subsection{Photoionization Calculations}
\label{subsec:photoionization_calculations}

Autoionizing states give rise to peaks with Fano line profiles~\cite{fano_1961} in the photoionization cross section. 
The line widths~$\Gamma$, and consequently, the lifetimes $\tau = \hbar / \Gamma$ of these resonant states can be obtained by calculating the cross sections, followed by fitting the peaks by the Fano formula~\cite{fano_1961} $a(q+\epsilon)^2/(1+\epsilon^2)+b$
with $\epsilon=(E-E_R)/(\Gamma/2)$, where $E$ is photon or photoelectron energy, $E_R$ is the resonant position, $q$ controls the line shape, $a$ describes the magnitude
of the cross section, and we assume a constant background $b$.

The cross section for a photon-induced single ionization of an atom or molecule initially in a state
$\Psi_i$ is, in the dipole approximation, proportional to the modulus squared of the dipole matrix 
element~\cite{lucchese_1982}
\begin{equation}
    \label{eq:xsec}
    \sigma(E) \propto \big\vert \big\langle\Psi_{f,\mathbf{k}}^{(-)}\big\vert \hat{\mu}\big\vert\Psi_i\big\rangle\big\vert^2\:,
\end{equation}
where the final state $\Psi_{f,\mathbf{k}}^{(-)}$ is a scattering wave function describing the
electron-cation complex with outgoing photoelectron with momentum~$\mathbf{k}$.
This $N$-electron wave function is written as a close-coupling expansion
\begin{equation}
    \label{eq:cc_expansion}
    \Psi_{f,\mathbf{k}}^{(-)}(\mathbf{r}^N) = \mathcal{A}\sum_\alpha  \Phi_\alpha(\mathbf{r}^{N-1})\chi_\alpha(\mathbf{r})\:,
\end{equation}
where $\mathbf{r}$ are electron coordinates,
$\Phi_\alpha$ are $(N-1)$-electron wave functions of bound ion states, the one-electron wave functions $\chi_\alpha$ describe the
outgoing photoelectron in channels~$\alpha$, and $\mathcal{A}$ is the antisymmetrizing operator. 

To compute the multichannel scattering wave functions $\Psi_{f,\mathbf{k}}^{(-)}$, the Schwinger variational
method was used as implemented in the \texttt{MCSCI} suite of codes designed for treating photoionization of linear
molecules.~\cite{Stratmann_schwinger1995,Stratmann_schwinger1996, botting_1997}
In this approach, one-electron wave functions for fixed energy were represented using
a single-center partial-wave expansion with $L_\mathrm{max}=16$ and with radial
parts on a grid up to 100~bohr. The initial $\Psi_i$ and ion $\Phi_\alpha$ wave functions were constructed using restricted or complete
active space (RAS or CAS) configuration interaction (CI) method from atomic orbitals obtained using the \texttt{MOLPRO} 
quantum chemistry package.~\cite{molpro_2012, molpro_2020, molpro_version_2022d3} 
It is not straightforward to optimize the $5d$ and $6d$ orbitals for the $4d^{-1}n\ell$ states without breaking the symmetry in \texttt{MOLPRO}
since the calculations are restricted to the D$_\mathrm{2h}$ group. Therefore, ion orbitals were used from a Hartree-Fock (HF) calculation for the $^2S\,(5s^{-1})$ state of Xe$^+$ calculated with relativistic
aug-cc-pVQZ-DK3 basis set~\cite{Bross2013} without $f$ and $g$ functions, but further augmented by five $s$, five $p$, and five $d$ diffuse Rydberg
functions.\cite{kaufmann1989}
For some photoionization calculations, HF orbitals for the Xe ground state computed with the standard
aug-cc-pVTZ-DK3 basis set were also used .~\cite{Bross2013} 
In Section~\ref{sunbsec:photoion_decay}, these orbitals are referred to as the ion and neutral orbitals.
All \texttt{MOLPRO} calculations used the Douglas-Kroll-Hess relativistic one-electron integrals.~\cite{reiher_2004, reiher_2004b}

\subsection{FWM Simulations}
\label{subsec:fwm_simulations}

To simulate the measured FWM signal, a TDSE-based multi-emitter calculation was employed.
The FWM signal originates from the far-field interference of the emission of multiple emitters under the influence of the transient NIR grating in the focus.
The transient NIR grating and its periodicity are defined by the noncollinear angles of the two NIR beams in respect to each other.
Hence, the FWM signal can be simulated by solving the TDSE for a one-dimensional chain of emitters orthogonal to the XUV beam and in the plane defined by all input beams.\cite{mi2021method}
Due to the noncollinear angle of the FWM experiment (see Fig.~\ref{fig:experimental_fwm_geometry}) and hence tilted NIR wavefronts, every emitter is under the influence of a different effective NIR field consisting of the coherent sum of NIR1 and NIR2, which acts as an input for the time-dependent Hamiltonian of the TDSE. 
The bright and dark states together with the ground state act as a basis for the few-level TDSE, which is solved with a split-step method.~\cite{bandrauk1993exponential}
Here, the electronic state energies, the transition dipole moments, and the lifetimes of the included states are parameters of the model.
The resulting excited electronic state wave functions of each emitter are used to calculate the time-dependent dipole moment, which gives rise to the near-field complex absorbance \cite{gaarde2011transient} and therefore the spatially resolved emitted electric field.
The resulting near-field signals are then propagated to the far field via Fraunhofer diffraction.
This FWM simulation approach is similar to the FWM simulations by Mi~\textit{et al.}\cite{mi2021method} used to predict FWM signals in doubly-excited states in helium, which were recently measured by the authors.\cite{rupprecht_extracting_2024}

\section{Results and Discussion}
\label{sec:resultsanddiscussion}

\begin{figure}
\includegraphics{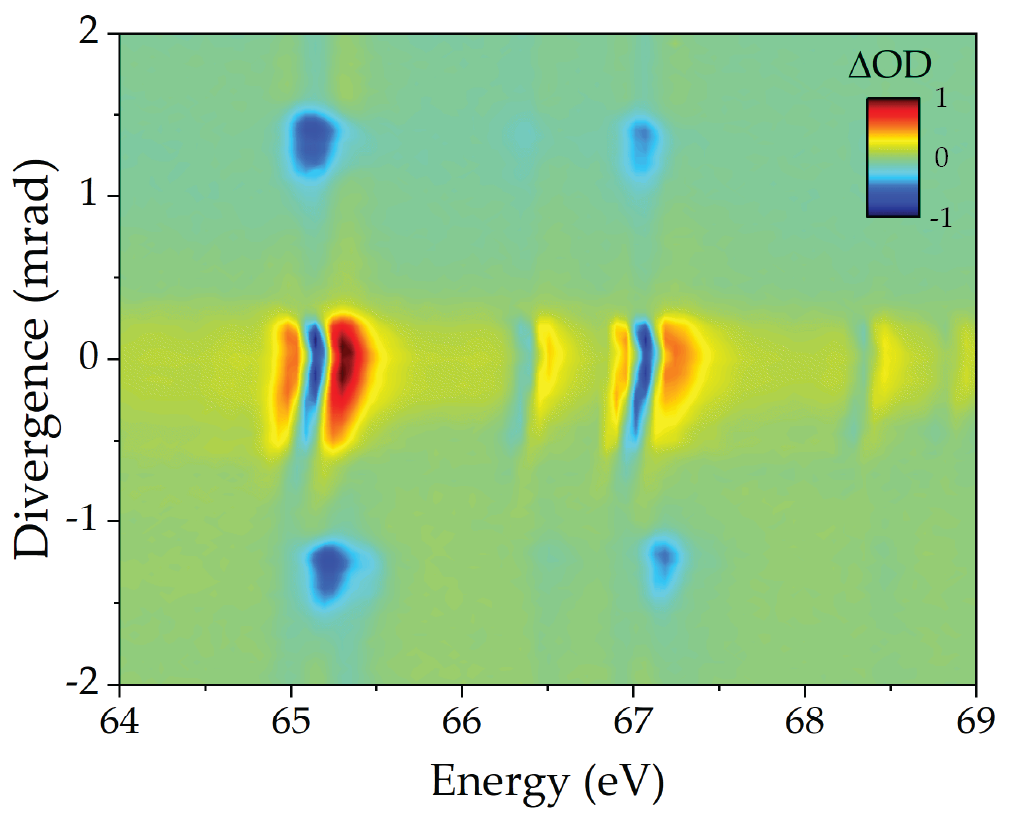}
\caption{\label{fig:2d camera image}
         XUV-spectrograph camera image of the FWM signal in xenon at temporal overlap ($\tau_1 = \tau_2 = 0)$. Absorption (red) occurs at a divergence of 0 mrad, while FWM emission (blue) occurs at a divergence of approx. $\pm$1.3 mrad. The states observed in this spectrum are $4d^{-1}_{5/2}6p$ at 65.11 eV, $4d^{-1}_{5/2}7p$ at 66.38 eV, $4d^{-1}_{3/2}6p$ at 67.04 eV, and $4d^{-1}_{3/2}7p$ at 68.35 eV.\cite{kolbasova_probing_2021}
}
\end{figure}

The XUV-spectrograph camera image in Fig.~\ref{fig:2d camera image} depicts the FWM signals at spatial and temporal overlap of all pulses in the target cell. The \textit{x}-axis, or photon energy in eV, is generated by spectrally dispersing the light with the XUV grating. The \textit{y}-axis, or divergence in mrad, quantifies the degree of FWM signal separation from the transmitted XUV signal, and arises from the phase-matching process with a noncollinear beam geometry [Eq.~\eqref{eqn:divergence_angle}]. The color scale is given in differential absorbance ($\Delta$OD), where absorption is indicated by positive $\Delta$OD signals (red) while emission is indicated by negative $\Delta$OD signals (blue). 
In Fig.~\ref{fig:2d camera image}, the differential static absorption spectrum is defined at 0 mrad and the FWM emission signals arise at approximately $\pm$1.3 mrad. All FWM signals appear as emission.


\subsection{Measured decay lifetimes}
\label{subsec:measured decay lifetimes}

\begin{figure*}[t]
\includegraphics{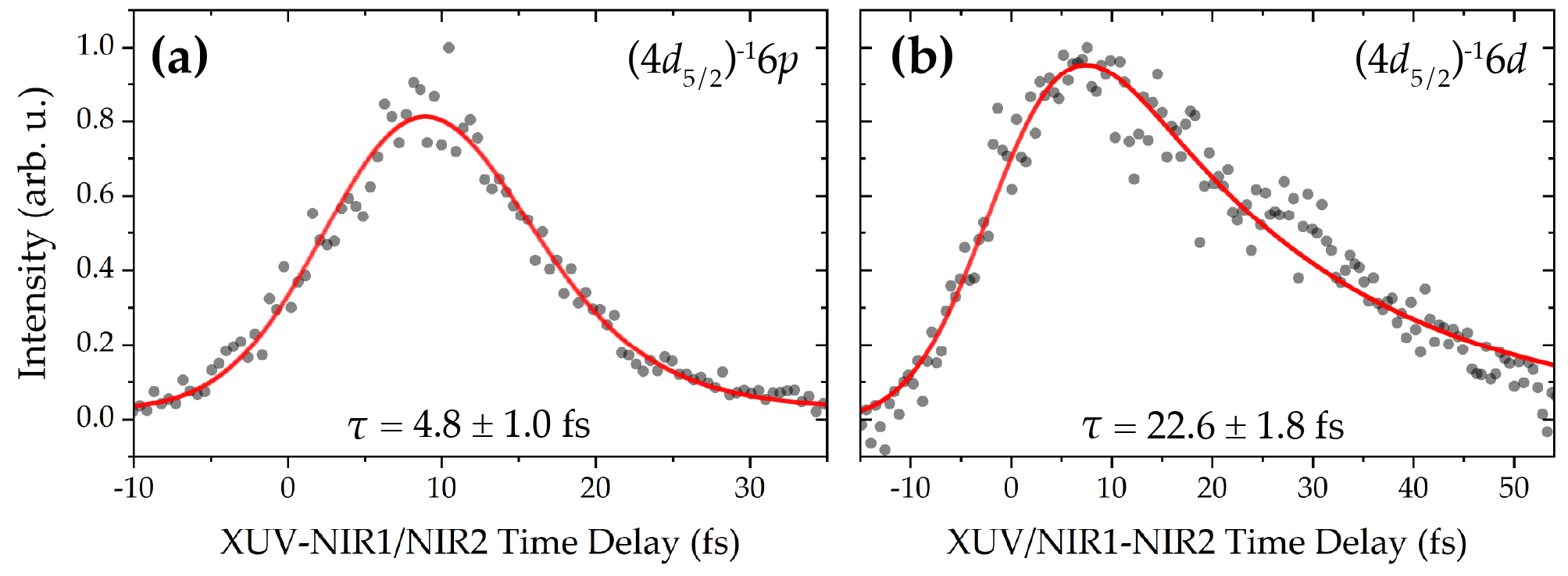}
\caption{\label{fig:lineouts}
         Time-resolved FWM lineouts of the lower FWM emission intensity of the (a) $4d^{-1}_{5/2}6p$ bright state and (b) $4d^{-1}_{5/2}6d$ dark state. The experimental values (gray circles) are the average of the FWM signature over its full spectral width. The fit (red line) is a convolution of a Gaussian function and a monoexponential decay function in order to extract the core-excited state decay lifetimes in the perturbative limit. The resulting decay lifetimes for the (a) $4d^{-1}_{5/2}6p$ bright state and (b) $4d^{-1}_{5/2}6d$ dark state are $4.8 \pm 1.0$ fs and $22.6 \pm 1.8$ fs, respectively, reported with 95\% confidence.}
\end{figure*}

One representative bright-state scan ($4d^{-1}_{5/2}6p$) and one representative dark-state scan ($4d^{-1}_{5/2}6d$) are presented in Fig.~\ref{fig:lineouts}(a) and \ref{fig:lineouts}(b). All other scans ($4d^{-1}_{5/2}6s$, $4d^{-1}_{3/2}6s$ $4d^{-1}_{3/2}6p$, $4d^{-1}_{3/2}6d$) can be found in SM Fig.~2. Using a vertically movable camera mount, the absorption signals seen in Fig.~\ref{fig:2d camera image} at 0~mrad were masked to enhance the signal-to-noise ratio of either the upper or the lower FWM emission signals by minimizing scattered background light. To obtain the time-resolved FWM lineouts in Fig.~\ref{fig:lineouts}, the spectral axis was binned and averaged over the full spectral width of the FWM signature at each time delay (0.4 fs steps). This process yields the experimental data (gray circles) that were fitted to a convolution of a Gaussian function and a monoexponential decay function (red line), which returns the core-excited state decay lifetimes.
Such a monoexponential fit reflects the expected population decay in the perturbative limit.

\begin{table}
    \caption{\label{tab:lifetimes}
    Summary of the $4d^{-1}$ core-excited state decay lifetimes in xenon extracted by fitting the experimental FWM data with a monoexponential decay model. For the time-resolved FWM signals not provided in Fig.~\ref{fig:lineouts}, see SM Fig.~2.}
    \begin{threeparttable}
    \begin{ruledtabular}
    \begin{tabular}{lcccc}
        state & state character & FWM lifetime & NIR intensity  \\
              &                 & (fs)         & (W/cm$^2$)     \\
        \colrule\\[-0.9em]
        $4d^{-1}_{5/2}6s$   & dark   & 17.6 $\pm$ 1.1 & $3.7 \times 10^{12}$ \\[0.2em]
        $4d^{-1}_{3/2}6s$   & dark   & 18.4 $\pm$ 1.6 & $3.7 \times 10^{12}$ \\[0.2em]
        $4d^{-1}_{5/2}6p$   & bright & 4.8 $\pm$ 1.0  & $4.4 \times 10^{12}$ \\[0.2em]
        $4d^{-1}_{3/2}6p$   & bright & 5.8 $\pm$ 1.3  & $4.4 \times 10^{12}$ \\[0.2em]
        $4d^{-1}_{5/2}6d$   & dark   & 22.6 $\pm$ 1.8 & $4.0 \times 10^{12}$ \\[0.2em]
        $4d^{-1}_{3/2}6d$   & dark   & 21.3 $\pm$ 3.1 & $4.0 \times 10^{12}$ \\[-0.2em]               
    \end{tabular}
    \end{ruledtabular}
    \end{threeparttable}
\end{table}

Fig.~\ref{fig:lineouts}(a) shows the temporal evolution of the lower FWM signal in a bright-state scan configuration, which measures the decay lifetime of the $4d^{-1}_{5/2}6p$ bright state. The fit of this profile yielded a decay lifetime of 4.8 $\pm$ 1.0 fs, which is in excellent agreement with all previous synchrotron linewidth measurements of the $4d^{-1}_{5/2}6p$ state (see SM Table~I). However, a different result occurs for the dark-state FWM signature in Fig.~\ref{fig:lineouts}(b). There, the extracted decay lifetime for the $4d^{-1}_{5/2}6d$ dark state is found to be 22.6 $\pm$ 1.8~fs. Interestingly, a $\sim20$~fs fitted lifetime was found for all the dark-states, revealing an apparent lifetime dependence on state character: with the monoexponential decay model, the $4d^{-1}_{\{5/2,3/2\}}6p$ bright states have autoionizing decay lifetimes of $\sim$5~fs while the $4d^{-1}_{\{5/2,3/2\}}6s$ and $4d^{-1}_{\{5/2,3/2\}}6d$ dark states all exhibit decay lifetimes in the 20 fs range (see Table~\ref{tab:lifetimes}). 

While the FWM experimental results presented in Fig.~\ref{fig:lineouts} support the established idea that the $4d^{-1}_{5/2}6p$ and $4d^{-1}_{3/2}6p$ bright states decay via spectator channels in approximately 5 fs, they do not immediately clarify whether the $4d^{-1}_{5/2}6s$, $4d^{-1}_{3/2}6s$, $4d^{-1}_{5/2}6d$, and $4d^{-1}_{3/2}6d$ dark states also decay via the same spectator channels due to their longer decay lifetimes. In the next section, we discuss photoionization calculations that explore the decay mechanisms of these states.

\subsection{Theoretical Photoionization Decay Channels}
\label{sunbsec:photoion_decay}

A series of photoionization calculations were performed in order to extract the line widths for the decay of $4d^{-1}n\ell$ 
states into participator and spectator ion channels. 
The calculations will show that the spectator channels dominate in the overall decay over the participator channels for all $4d^{-1}n\ell$ states with linewidths that are all very similar, giving expected decay times of 5--6~fs for both bright and dark states of $4d^{-1}n\ell$ character.
All of the calculations discussed here used LS coupling for the atomic states, that is,
the spin-orbit interaction is not taken into account but both singlet and triplet spin states were considered.
In order for an autoionizing state of interest to affect the photoionization cross section, the parent ion state of this
resonance\footnote[2]{The parent ion state is such an ion state that if an additional electron is attached,
the resulting neutral state describes the principal configuration of the resonance.}
must be included in the close-coupling expansion [Eq.~\eqref{eq:cc_expansion}] and the resonance must have a symmetry that allows a one-photon transition from 
the initial state. In the case of the $4d^{-1}n\ell$ resonances, the parent ion state is $^2D\,(4d^{-1})$. 

The principal configuration $4d^{-1}6p$ for the bright state leads to three atomic states $^1P^o$, $^1D^o$, and $^1F^o$, but only the 
$^1P^o$ is accessible by one-photon transition from the ground state $^1S$. 
The dark states are dipole forbidden with respect to the ground state, but any
initial state of $^1P^o$ symmetry can be used to access the dark states since the line widths are independent of the initial state.
There is only one dark state $^1D$ corresponding to the $4d^{-1}6s$ configuration. For $4d^{-1}nd$ configurations
we can get $^1S$, $^1P$, $^1D$, $^1F$, and $^1G$ dark states, but only $^1S$ and $^1D$ can be accessed by a two-photon transition 
in the experiment. The $^1P$ state cannot be populated from the $^1P^o$ bright state 
because the NIR and XUV pulses have the same linear polarization. In what follows, calculations are presented for the decay widths
of $^1P^o\,(4d^{-1}6p)$, $^1D\,(4d^{-1}6s)$, and $^1S$, $^1P$, $^1D$ states with $4d^{-1}nd$, $n=5,6$. Triplet spin states $^3P^o\,(4d^{-1}6p)$ and $^3D\,(4d^{-1}6s)$ are also considered.

\begin{figure*}
\includegraphics{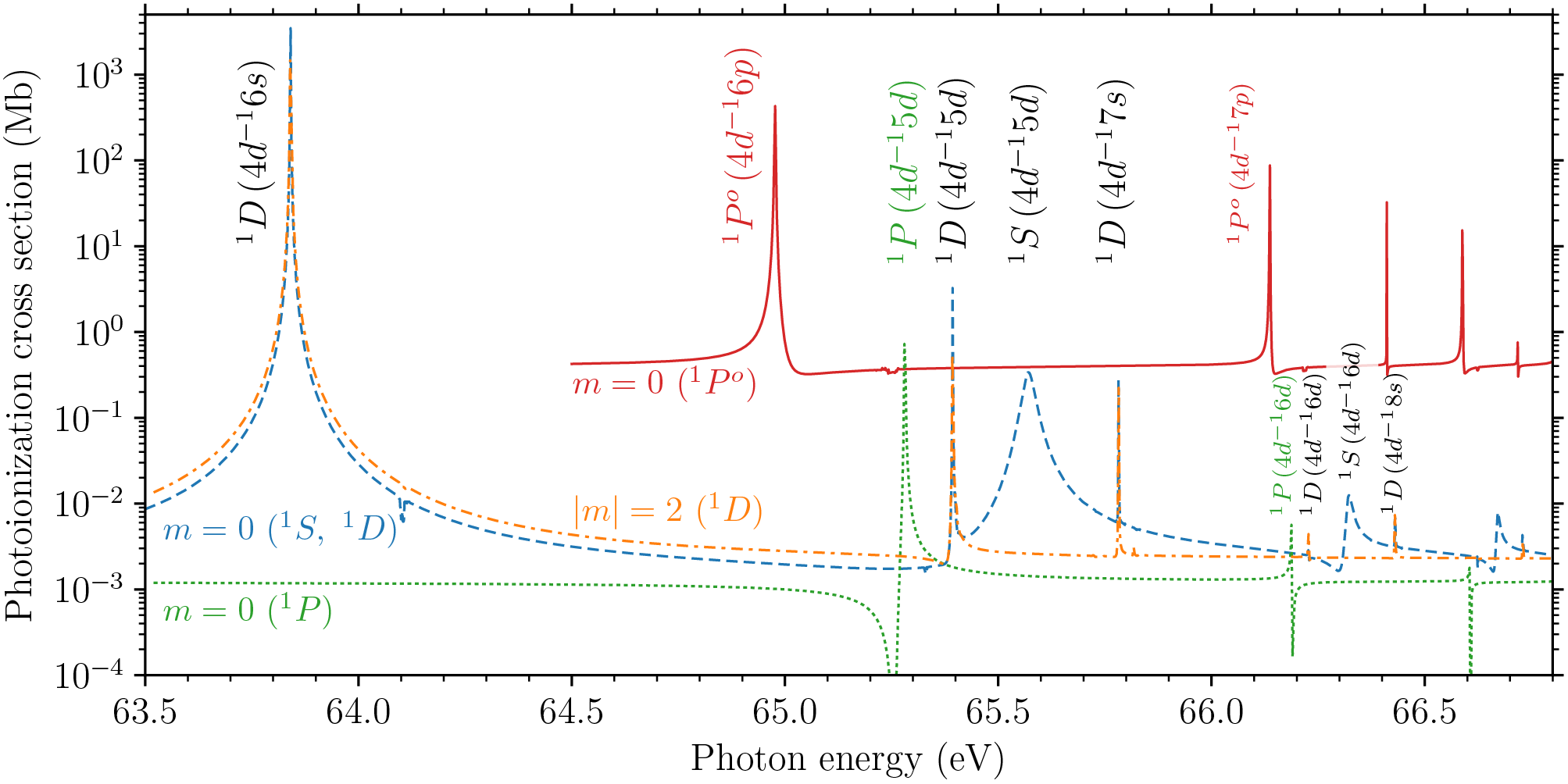}
\caption{\label{fig:xsec}
         Calculated cross sections for photoionization of xenon initially either in the ground state $^1S$ (red solid line),
         or in the first excited state $^1P^o\,(5p^{-1}6s)$ for the dark states (other lines). The calculations considered only the LS coupling for atomic states.
         The $m = 0$ component of the $^1P^o$ total symmetry (red solid) reveals the Rydberg series of the bright states $4d^{-1}np$. 
         For the $4d^{-1}ns$ and $4d^{-1}nd$ dark states we show $m=0$ (blue dashed) for $^1S$ or $^1D$ (not separated in our calculation),
         $m=0$ (green dotted) for $^1P$, and $m=2$ (orange dash-dotted) for $^1D$, see the text for details.
         The close-coupling expansion included only the participator channels $^2P^o\,(5p^{-1})$, $^2S\,(5s^{-1})$, and also $^2D\,(4d^{-1})$, which is 
         the parent state for all the autoionizing states $4d^{-1}n\ell$. The calculation was performed with RAS $4d5s5p5d6s6p$ and HF orbitals
         for Xe$^+$ $^2S\,(5s^{-1})$.}
\end{figure*}

To make the discussion easier to follow, an overall picture is shown in Fig.~\ref{fig:xsec} with Fano line profiles for all $4d^{-1}n\ell$ states. These calculated line profiles include only the decay into the participator channels $^2P^o\,(5p^{-1})$ and $^2S\,(5s^{-1})$, leading to very narrow lines. Later in this section, the quantitative contributions of participator and spectator channels for each $4d^{-1}n\ell$ state of interest are discussed in detail.
In this work, the \texttt{MCSCI} photoionization code~\cite{Stratmann_schwinger1995,Stratmann_schwinger1996, botting_1997} designed for linear molecules was used, see Sec.~\ref{subsec:photoionization_calculations}, which implements the D$_{\infty\mathrm{h}}$ point group. 
In D$_{\infty\mathrm{h}}$, $P$ and $D$ atomic symmetries are no longer irreducible and
have $\Sigma$, $\Pi$ and $\Sigma$, $\Pi$, $\Delta$ components that correspond to $\vert m\vert =0,1$ and $\vert m\vert=0,1,2$ components of the atomic states, respectively.
The bright states $4d^{-1}np$ appear in the cross section for photoionization from the ground state and Fig.~\ref{fig:xsec} shows 
the $^1\Sigma_u^+$ component of the total symmetry that corresponds to the $m = 0$ component of $^1P^o$. 
Note that the line widths are independent of the component of the total symmetry.
To reveal the dark states, the first excited state $^1P^o\,(5p^{-1}6s)$ was used as the initial state. Calculating the $^1\Sigma_g^+$ ($m=0$ for
$^1S$ and $^1D$), $^1\Sigma_g^-$ ($m=0$ for $^1P$), and $^1\Delta_g$ ($\vert m\vert =2$ for $^1D$)
components distinguishes the $^1S$, $^1P$, and $^1D$ states with $4d^{-1}nd$ configurations, see Fig.~\ref{fig:xsec}.

The CI method with four active spaces was used to calculate the initial and ion states for the photoionization calculations (see also Sec.~\ref{subsec:photoionization_calculations}):
two restricted active spaces (RAS) $4d5s5p5d6s6p$ and $4d5s5p5d6d$ in combination with the ion orbitals and two small complete active spaces (CAS) $4d5s5p6s$ and $4d5s5p6p$ with the neutral orbitals. Here, for example, $4d5s5p6p$ means that the active space consisted of $4d$, $5s$, $5p$, and $6p$ orbitals while the remaining lower-lying orbitals were kept frozen. In the RAS cases, 
the excitations were restricted to three- and two-electron excitations to the virtual orbitals ($5d$, $6s$, $6p$ or $5d$, $6d$)  
for the initial and ion states, respectively. The RAS $4d5s5p5d6s6p$ was used for the cross sections shown in Fig.~\ref{fig:xsec}.

Our nonrelativistic positions of the $4d^{-1}n\ell$ states 
are in good qualitative agreement with relativistic energies of Kolbasova \textit{et al.}~\cite{kolbasova_probing_2021}
The $4d^{-1}6s$ and $4d^{-1}6d$ dark states are positioned 1.0--1.5~eV below and above the bright state, respectively, and
the $4d^{-1}5d$ dark states lie $\sim$0.5~eV above the bright state; compare Figs.~\ref{fig:xenon_energy_level_diagram} and~\ref{fig:xsec}, see also Table~II in the SM.

The calculations shown in Fig.~\ref{fig:xsec} considered only the participator channels $^2P^o\,(5p^{-1})$ and $^2S\,(5s^{-1})$
as possible decay pathways for the autoionizing states. However, the bright states and most of the dark states do not strongly decay via the participator channels.
For each $4d^{-1}n\ell$ autoionizing state of interest, spectator decay channels with principal configurations of the ion states $5p^{-2}n\ell$, $5s^{-1}5p^{-1}n\ell$, and $5s^{-2}n\ell$ were further considered.
In theory, all decay channels should be included in the close-coupling expansion [Eq.~\eqref{eq:cc_expansion}] 
at the same time. However, such a calculation would be very costly or unfeasible. Therefore, the cross sections were calculated 
for each decay channel separately. Because the autoionizing states of interest are narrow, the total decay width
can be well-approximated by the sum of the widths for individual decay channels.~\cite{mccurdy_1979,botting_1997}

Considering the participator and spectator decay channels described above, the width of the $^1P^o\,(4d^{-1}6p)$ bright state was calculated to be
79~meV using the RAS $4d5s5p5d6s6p$ with the ion orbitals and 56~meV using the CAS $4d5s5p6p$ with the neutral orbitals, see Table~\ref{tab:widths_bright_state}.
Using the CAS, the width of the triplet spin variant of the bright state is 59~meV, that is, basically identical 
to the singlet variant. For the triplet calculation, the lowest $^3S$ state was used as the initial state.
The $\sim$20 meV difference in the widths between the RAS and CAS calculations is primarily given by 
the very limited complete active space and better orbitals optimized for the Xe$^+$~$^2S\,(5s^{-1})$ used in the calculation with the restricted space. 
Synchrotron experiments~\cite{heimann_1987, becker_1989} found that the bright state significantly (30--40~\%) decays via
ejection of multiple electrons (shake off). Double- and triple-ionization thresholds of xenon are 33.1~eV and 64.4~eV,
respectively.~\cite{dutil_1980} The calculations did not include any shake-off channels, but assuming 30--40~\% contribution 
for these channels, the total width would be 110--130 meV (lifetime of 6--5~fs) for both singlet and triplet bright states,
which is in good agreement with experiments.~\cite{ederer_photoabsorption_1975, king_investigation_1977, masui_new_1995, sairanen_high-resolution_1996, zhang_fast-electron-impact_2015, aksela_correlation_1995, aksela_electron_1994}

\begin{table}[t!]
    \caption{\label{tab:widths_bright_state}
    Calculated line widths $\Gamma$ (in meV) for the decays of the $^1P^o$ and $^3P^o$ bright states $4d^{-1}6p$ into participator channels $5p^{-1}$ and $5s^{-1}$ and various spectator channels (the rest of channels).
    The calculations were performed with two active spaces. In the case of RAS, only the singlet variant was calculated. 
    The first row reports the width from a calculation with both participator channels included at the same time.
    The last row gives the lifetime from the total width. The dash --- means that the width was not calculated.
    Note that the total widths do not include any shake-off channels; see the text for estimates.}
    \begin{threeparttable}
    \begin{ruledtabular}
    \begin{tabular}{lccc}
        \multirow{2}{*}{\makecell{principal \\ configuration}} & \multirow{2}{*}{\makecell{RAS $4d5s5p5d6s6p$\tnote{a} \\ $\Gamma(^1P^o)$}} & \multicolumn{2}{c}{CAS $4d5s5p6p$\tnote{b}} \\ & & $\Gamma(^1P^o)$ & $\Gamma(^3P^o)$ \\ 
        \colrule\\[-0.9em]
        $5p^{-1}$, $5s^{-1}$ & 2.0 & 1.4 & 0.4 \\[0.2em] 
        $5p^{-1}$ & 2.0 & --- & --- \\[0.2em] 
        $5s^{-1}$ & 0.2 & --- & --- \\[0.2em] 
        $5p^{-2}6p$ & 41.4 & 26.8 & 29.2 \\[0.2em] 
        $5s^{-1}5p^{-1}6p$ & 10.0 & 5.8 & 6.3 \\[0.2em] 
        $5p^{-3}5d\,6p$ & 0.6 & --- & --- \\[0.2em] 
        $5s^{-2}6p$ & 24.5 & 21.9 & 22.6 \\[0.2em] 
        \colrule\\[-0.9em] 
        total width & 78.6 & 55.9 & 58.5 \\[0.2em] 
        \colrule\\[-0.9em] 
        lifetime (fs) & 8.4 & 11.8 & 11.3 \\
    \end{tabular}
    \end{ruledtabular}
    \begin{tablenotes}\footnotesize
        \item[a] with HF orbitals for Xe$^+$ $^2S\,(5s^{-1})$
        \item[b] with HF orbitals for Xe ground state
    \end{tablenotes}
    \end{threeparttable}
\end{table}

Table~\ref{tab:widths_bright_state} further shows the contribution of considered ion channels grouped based on
the principal configuration of the residual ion. The widths for each individual ion state are given in Tables~III and~IV in the SM.
The $5p^{-2}6p$ spectator channel dominates, which agrees with the experiment of Aksela \textit{et al.}~\cite{aksela_electron_1994}
From analysis of their synchrotron experiment, Becker \textit{et al.}~\cite{becker_1989}
attributed up to about 20\% of the overall decay
to $5p^{-3}5d6p$ spectator channels. 51 ion states with this principal configuration were found lying in the 41 to 47~eV region but the calculations
give the combined width of only 0.6~meV for all these channels. The sum of the widths in the second and third rows (participator channels considered separately) of Table~\ref{tab:widths_bright_state} should be equal to the width from the first row (both participators channels
included at the same time).
The difference gives an estimate of the error in the determination of the partial widths, which is in general $\sim$0.1--1.0~meV. 

The calculated total width for the $^1D\,(4d^{-1}6s)$ dark state is 
77~meV using the RAS $4d5s5p5d6s6p$ and 63~meV using the CAS $4d5s5p6p$ (see Tables~V, VI, and~VII in the SM for details). 
The width for the triplet spin state, again calculated only with the CAS space, is 62~meV. No significant difference was observed in the total widths or contributions of decay channels in comparison to the bright state.
Since the $5p^{-3}5dn\ell$ channels had a negligible contribution to the decay of the bright state (see above), they were not considered for any dark state. 
Considering again $\sim$30~\% contribution of 
shake-off channels, the total width from the RAS calculation would be 110~meV, which is in good agreement
with the experimental width 102~meV of Zhang \textit{et al.}~\cite{zhang_fast-electron-impact_2015} for the $4d^{-1}_{5/2}6s$ state.

The restricted active space $4d5s5p5d6s6p$ cannot be used to calculate the contributions of spectator channels for the $4d^{-1}6d$
dark states because the $6d$ orbitals are missing in the active space to construct the final ion states. Therefore, another restricted space, RAS $4d5s5p5d6d$, was used to obtain ion channels for the $4d^{-1}6d$ states. Both these spaces were used for the $4d^{-1}5d$ states in order to be able to judge the effect of omitting the $6s$ and $6p$ orbitals.

Using the RAS $4d5s5p5d6s6p$, the calculated decay widths of $^1S$, $^1P$, and $^1D$ $4d^{-1}5d$ dark states are 109, 77, and 77~meV, respectively. The RAS $4d5s5p5d6d$ gives $\sim$10~\% smaller total widths
(Tables~VIII and~IX in the SM). The $^1P$ and $^1D$ states decay almost identically to 
the $4d^{-1}6p$ bright and $4d^{-1}6s$ dark states. In contrast, the $^1S\,(4d^{-1}5d)$ state decays 
quite significantly with the width of 32~meV via the two participators channels, with dominant decay (25.5~meV)
to the ground ion state $^2P^o\,(5p^{-1})$.
The widths for the spectator decay of $^1S$, which still dominate the overall decay, are the same as for other states, which results in the total
width without shake-off channels of 109~meV and lifetime of 6~fs. Taking into account a $\sim$30~\% contribution of shake-off states, the total width is $\sim$160~meV (lifetime $\sim$4~fs) for $^1S$ and $\sim$100~meV (lifetime $\sim$6.6~fs) 
for the $^1P$ and $^1D$ states with the $4d^{-1}5d$ character.
 
Finally, the widths for $^1S$, $^1P$, and $^1D$ $4d^{-1}6d$ states were calculated with the restricted active space $4d5s5p5d6d$. The calculation gives the same decay widths of 72 and 70~meV for $^1P$ and $^1D$, respectively, as for the 
analogous $4d^{-1}5d$ states with the same active space (Tables~X and~XI in the SM).
The width of the $^1S\,(4d^{-1}6d)$ state for the decay into the participator channels is about half (16.3~meV) in comparison to the $^1S\,(4d^{-1}5d)$.
Further, the widths for the decay of $^1S\,(4d^{-1}nd)$ states for $n=5,\ldots,9$ into the participator channels are
 32.2, 15.4, 7.9, 4.6, and 2.9~meV, which follow the $n^3$ behavior as shown by Fano and Cooper.~\cite{fano_line_1965}
The triplet spin states for $4d^{-1}nd$ states were not considered
but based on the $4d^{-1}6p$ and $4d^{-1}6s$ states it is expected that they decay in the same way as the corresponding singlet states.

To summarize our nonrelativistic photoionization calculations, lifetimes of 6--9~fs were obtained for the decay of $4d^{-1}n\ell$ 
autoionizing states via single-ionization channels (without shake off). No significant differences in the decay of the $4d^{-1}n\ell$ ($n\ell=6s$, $5d$, $6d$)
dark states were found in comparison to the $4d^{-1}6p$ bright state. The decay is dominated by $5p^{-2}n\ell$ spectator channels, followed by
the $5s^{-2}n\ell$ spectator channels. The only exception is that $4d^{-1}nd$ states of the total $S$ symmetry have a stronger contribution of the
participator channels with widths of 32 and 16~meV for $n=5$ and 6, respectively, compared to widths of 0.5--2~meV for other $4d^{-1}n\ell$ states. 
Taking into account $\sim$30~\% contribution of the shake-off decay channels,~\cite{heimann_1987, becker_1989}
the $4d^{-1}n\ell$ states have lifetimes of 4--6~fs. 
Furthermore, almost no difference between the singlet and triplet spin states was observed. 
As a result, the xenon states of $4d^{-1}n\ell$ character in the $jj$ coupling will have basically the same lifetimes since these states
are given by linear combinations of the states within the LS coupling with the corresponding $J$ quantum number.

It is not surprising that all $4d^{-1}n\ell$ states have approximately the same lifetime since the electron is excited
from the $4d$ orbital to Rydberg-like orbitals, leading to a weak interaction of the excited electron
with the core hole. 
Finding no reason that the dark states of the $4d^{-1}n\ell$ character could have longer lifetimes, as observed experimentally, an alternative kinetic-pumping mechanism is considered next for the observed longer timescales for the dark states by FWM.

\subsection{Simulated FWM Signal}
\label{sec:FWM_simulation:_results}

\begin{figure*}[t!]
\includegraphics{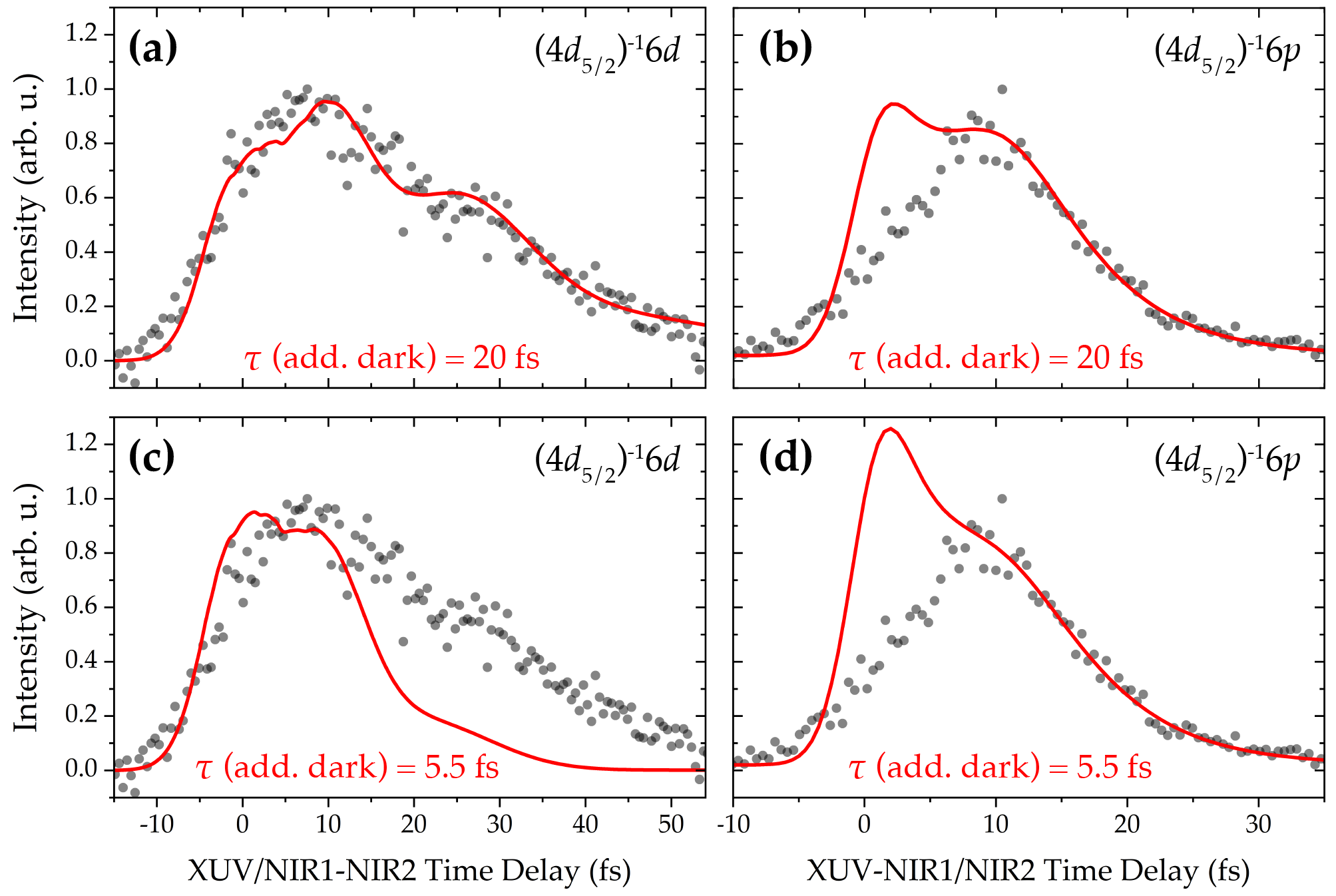}
\caption{\label{fig:sim_lineouts}
         Influence of dark-state lifetime on FWM TDSE simulations for the $4d^{-1}_{5/2}6p$ and $4d^{-1}_{5/2}6d$ states including an additional dark state located at 66.65~eV. The simulation result is shown as solid red line while the experimental data points are shown as black dots. (a) and (c) depict the $4d^{-1}_{5/2}6d$ dark-state scans simulated with a 20~fs and 5.5~fs additional dark-state lifetime, respectively. Here, only the longer lifetime simulation in (a) is in agreement with the experimental data. (b) and (d) show the $4d^{-1}_{5/2}6p$ bright-state scans simulated with a 20~fs and 5.5~fs additional dark-state lifetime, respectively. Here, the dark-state lifetime does not influence the decay dynamics.}
\end{figure*}

An alternative kinetic-pumping mechanism leading to the significantly longer FWM dark-state-scan signal considers other highly excited dark states of non-4d-core-hole character. 
Based on previous FWM analyses \cite{puskar_measuring_2023, rupprecht_extracting_2024}, optical  coupling to another nearby dark state of different configuration and longer lifetime could explain the longer lifetimes observed in the experimental dark-state scans.
Hence, to gain more insights into the physical cause of the extended signal of Fig.~\ref{fig:lineouts}(b), a few-level TDSE calculation (see also Sec.~\ref{subsec:fwm_simulations}) is employed: it includes the xenon ground state, the $4d^{-1}_{5/2}6p$ bright state, the $4d^{-1}_{5/2}6d$ dark state, and an additional dark state located at 66.65~eV. 
The electronic character of such a state is discussed in Sec.~\ref{subsec:darkState_decay}. 
A transition dipole matrix element of 2~atomic units (a.u.) for the $4d^{-1}_{5/2}6p \rightarrow 4d^{-1}_{5/2}6d$ transition is chosen, which is in agreement with both the values of Kolbasova \textit{et al.}\cite{kolbasova_probing_2021} as well as the \textit{ab-initio} calculations in Sec.~\ref{sunbsec:photoion_decay}.
In addition, the transition from the $4d^{-1}_{5/2}6p$ bright state to the additional dark state is chosen to be 0.88~a.u., which is weaker than the before-mentioned bright-to-dark state transition.
The beam geometry and pulse parameters are according to the experimental conditions: A 1.5$^{\circ}$ noncollinear angle, an NIR center wavelength of 800~nm, and NIR pulse durations of 7.4~fs and 8.9~fs for NIR1 and NIR2 (see SM Fig.~1), respectively, as measured with a home-built pulse-characterization apparatus.\cite{rupprecht_flexible_2023} 
Further details on the simulation are given in SM Sec.~II.\par
If a 5.5~fs lifetime for $4d$ core-hole states is used in combination with an additional dark-state lifetime of 20~fs, the resulting TDSE FWM simulation is in good agreement with the experimental $4d^{-1}_{5/2}6d$ dark-state scan [see Fig.~\ref{fig:sim_lineouts}(a)].
This simulation reproduces the decay structure (e.g. the plateau at 25~fs) that is not captured by the monoexponential fit in Fig.~\ref{fig:lineouts}(b).
\par
If all lifetimes are set to 5.5~fs (including the additional dark state), the simulated FWM signal of the $6d$ dark state significantly deviates from the experimentally observed decay profile [see Fig.~\ref{fig:sim_lineouts}(c)].
Furthermore, to investigate the influence of a longer-lived dark state on the extracted  $4d^{-1}_{5/2}6p$ bright state signal, the TDSE simulation was repeated for a bright-state-scan pulse sequence for both additional dark-state lifetimes of 20~fs and 5.5~fs, see Fig.~\ref{fig:sim_lineouts}(b) and Fig.~\ref{fig:sim_lineouts}(d), respectively.
As expected, the bright-state decay dynamics remain unaffected by the dark-state lifetime and capture the 5.5~fs $4d^{-1}_{5/2}6p$ lifetime.\par 
Both of the simulated bright-state scans show a Gaussian component at the temporal overlap that is not present in the measured data.
This feature has been observed both theoretically \cite{mi2021method} and experimentally \cite{rupprecht_extracting_2024} in other FWM studies and is linked to the enhancement of non-resonant FWM pathways in temporal overlap of the XUV and NIR1/NIR2 pulses. 

\begin{figure*}[t!]\textbf{}
\includegraphics{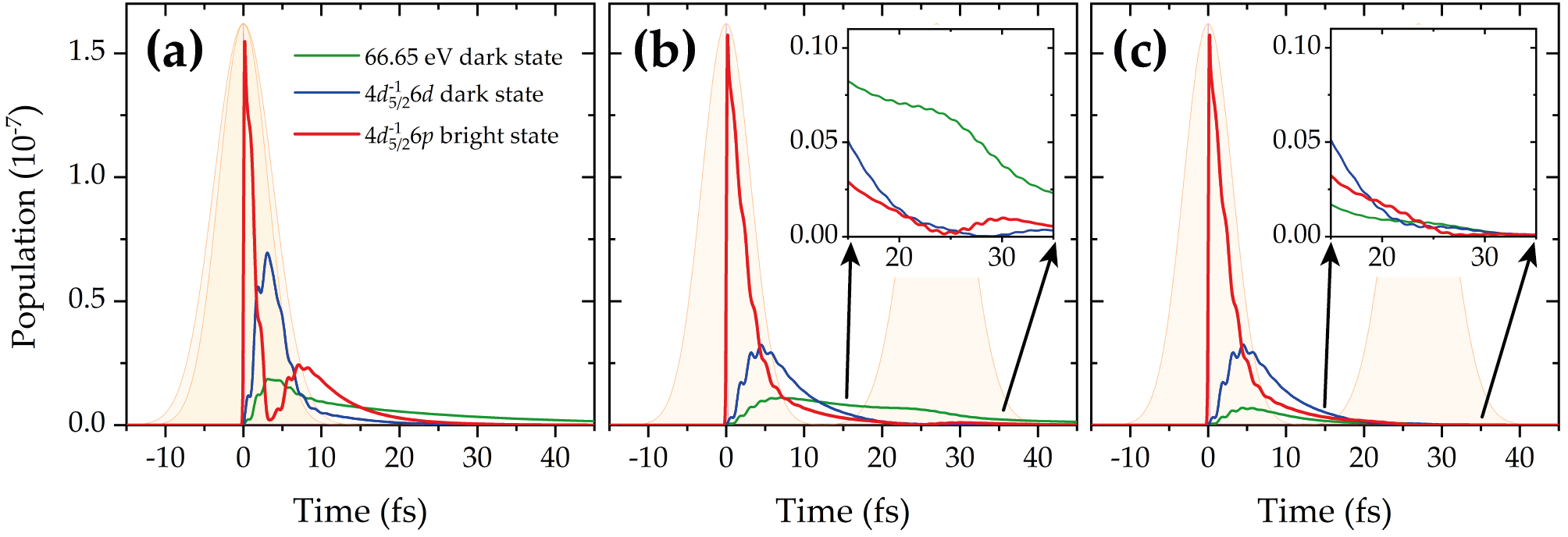}
\caption{\label{fig:population}
         Population dynamics from TDSE few-level simulation for a xenon atom centered within the emitter chain. The $4d^{-1}_{5/2}6p$ bright-state population is given as solid red line, the $4d^{-1}_{5/2}6d$ dark-state population as solid blue line, and the additional dark-state population (located at 66.65~eV) as solid green line. The XUV pulse is centered at 0~fs. The NIR1 and NIR2 intensity envelopes are shaded in orange. The population dynamics in temporal overlap (a), as well as for a dark-state scan time delay of 25~fs [(b) and (c)] are shown. The insets in (b) and (c) depict a zoomed-in view around the NIR2 pulse centered at 25~fs. In (a) and (b), the additional dark state has a 20~fs lifetime, while in (c) it has a lifetime of 5.5~fs.}
\end{figure*}

This coherent enhancement is evident in the population dynamics in overlap versus out of overlap as shown in Fig.~\ref{fig:population}(a) and Fig.~\ref{fig:population}(b), respectively.
While this TDSE model that only includes three excited states can simulate the decay dynamics well, its limited basis might be one reason why it deviates from the experimental data concerning this Gaussian coherence artifact. 
In addition, any difference in resonance energy position of the dark states, center photon energy, and transition dipole moments will have a proportionally larger effect on the coherence artifact in comparison to the decay dynamics. 
These simulations were performed for an NIR intensity of $9\times10^{11}$~W/cm$^2$, which is about four times lower than the estimated experimental intensities (see Table~\ref{tab:lifetimes}).
Notably, other FWM studies also had to considerably adapt the NIR intensity for respective simulations to reproduce experimental data.\cite{yanez2024non}
The experimental NIR intensity estimate in the target region is uncertain as it is deduced from the pulse measurements taken outside of the vacuum system, namely NIR power, imaged focus size, and temporal pulse characteristics. 
Taking these factors into consideration, the difference in the intensities for theory versus experiment is not surprising, especially as they are still within the same order of magnitude.\par
This simulation demonstrates that a lengthened FWM dark-state-scan signature necessitates the presence of an additional dark state that is significantly longer-lived than all calculated $4d$ core-hole states. 
Furthermore, due to the utilized NIR intensities, which were necessary to achieve a reasonable signal-to-noise ratio for the measured FWM experiments, strong-field effects come into play, which led to a deviation from the monoexponential FWM signal decay of perturbative NIR state couplings.
The exemplary population dynamics of an emitter in the middle of the emitter chain for an NIR intensity of $9\times10^{11}$~W/cm$^2$ shows considerable Rabi cycling in temporal overlap [see Fig.~\ref{fig:population}(a)].
This plot demonstrates that the high coupling strength induced by the rather large bright-to-dark state transition dipole moments\cite{kolbasova_probing_2021} in combination with high NIR peak intensities close to the TW regime lead to strong-field effects.
Hence, this simulation is particularly useful as intuitive FWM signal arguments for the perturbative regime are not valid anymore. 
The contribution from strong field effects is also highlighted by the qualitative temporal evolution of the simulated dark-state scan if all contributing states had a 5.5~fs lifetime. 
In the perturbative regime, this would lead to a monoexponential decay FWM signature. 
The simulation depicted in Fig.~\ref{fig:sim_lineouts}(c), though, shows a more complex structure. \par
To get a deeper insight into population dynamics that lead to the different FWM signatures for different lifetimes of the additional dark state, one can focus on a time-delay regime where they differ significantly, specifically the bump at 25~fs  in Fig.~\ref{fig:sim_lineouts}(a).
Figures~\ref{fig:population}(b) and \ref{fig:population}(c) depict the population dynamics for a 25~fs XUV/NIR1--NIR2 time delay if the additional dark-state lifetime was 20~fs and 5.5~fs, respectively.
The insets show a detailed view of the population dynamics while the NIR2 pulse is acting upon the xenon atom.
The FWM emission is ultimately governed by the $4d^{-1}_{5/2}6p$ bright-state population after interaction with all laser pulses, as only this state within the utilized basis can couple back to the ground state, which is necessary for XUV FWM emission. \par
Here, one can see that for the 20~fs lifetime case [Fig.~\ref{fig:population}(b)], the $4d^{-1}_{5/2}6p$ bright state is first depopulated, but afterwards repopulated from the additional dark state within the NIR2 pulse duration.
Hence, the resulting non-zero $4d^{-1}_{5/2}6p$ population after the NIR2 pulse can contribute to the respective FWM emission.
If the additional dark-state lifetime is 5.5~fs [Fig.~\ref{fig:population}(c)], however, barely any $4d^{-1}_{5/2}6p$ bright-state population is observed after the NIR2 pulse.
Therefore, the FWM emission will be much less intense at a time delay of 25~fs compared with the case of a longer dark-state lifetime, which agrees with the FWM simulations [Fig.~\ref{fig:sim_lineouts}(a) versus Fig.~\ref{fig:sim_lineouts}(c)].

\subsection{Configuration Character of Possible Dark-State Decay}
\label{subsec:darkState_decay}
While the TDSE simulations show the requirement of a nearby longer-lived dark state to replicate the experimental FWM signal, they do not identify which state that might be.
As previously discussed in Sec.~\ref{subsec:photoionization_calculations}, all $4d^{-1}n\ell$ core-hole states primarily decay via spectator channels and thus are expected to have a $\leq 6$~fs lifetime.
Therefore, the longer-lived additional dark state must be of a different character, specifically without a $4d$ core hole.
Using the RAS~CI method, about 20 autoionizing states of neutral xenon were found lying within a range from 63 to 67~eV, accessible by a one-photon transition from the bright state. Although most of these states have nearly zero transition dipole moments to the bright state, two states lying 0.38 and 0.46~eV above the calculated energy of the bright state with dominant configurations $5s^{0}5p^{5}5d^{2}6p^{1}$ and $5s^{0}5p^{5}5d^{1}6s^{1}6p^{1}$ have transition dipole moments of 0.5 and 0.2~a.u., respectively.
\par
While the lifetimes of these two states were not calculated due to convergence issues of the iterative Schwinger method, a longer lifetime of inner- and outer-valence holes in comparison to core 
holes is intuitively reasonable.\cite{fano_line_1965} Also, these two states are considerably more off-resonant to the NIR-transitions compared 
to the assumed energy of the additional dark state at 66.65~eV, which leads to the best agreement of the TDSE 
simulations with the measured data. However, the CI calculations presented here, which use a molecular code, are rather limited and do not 
include the spin-orbit interaction, which is expected to affect $4d$ core-hole states more than the states with only 
valence excitations. Thus, further calculations with relativistic atomic codes are desirable to identify the longer-lived 
dark states. \par
Possible candidates for longer-lived dark states with matching energies might be the different Rydberg series of these states converging to different excited ionic states. 
This might give rise to coupling to an ensemble or even a continuum of states as has been previously observed for FWM studies of autoionizing states in helium. \cite{rupprecht_extracting_2024}
Coupling to multiple states might also explain the difference in 0.88~a.u. transition dipole moment to the additional dark state in the TDSE in comparison to the $\leq 0.5$~a.u. dipole moments calculated for the states with mixed $5s^{0}5p^{5}5d^{2}6p^{1}$ and $5s^{0}5p^{5}5d^{1}6s^{1}6p^{1}$ characters. 
Here, the additional dark state in the FWM simulation acts as an effective dark state that includes the coupling to a whole dark-state manifold.
Furthermore, we see a similarly extended signature in the upper FWM signal that should be sensitive to the $4d^{-1}_{5/2}6s$ state lifetime in the absence of additional dark states.
Therefore, the appearance of elongated FWM signals in the upper and lower FWM dark-state-scan measurements is a further indication that there is a manifold of multi-electronic excited dark states in this energy region. 

\section{Conclusions}
\label{sec:conclusions}

In summary, attosecond noncollinear FWM spectroscopy was used to study the autoionization decay mechanisms and lifetimes of both the XUV-bright $4d^{-1}6p$ and XUV-dark $4d^{-1}6s$ and $4d^{-1}6d$ core-excited states. The initial hypothesis was that the $4d^{-1}6s$ and $4d^{-1}6d$ dark states would decay as spectator transitions akin the $4d^{-1}6p$ bright state, yielding natural decay lifetimes of approximately 6 fs for all $4d^{-1}6\ell$ states. However, while the experimental data recovered these 6~fs decay lifetimes for the $4d^{-1}6p$ bright states, decay lifetimes of $\sim$20~fs were measured in dark state scans that nominally probed the $4d^{-1}6s$ and $4d^{-1}6d$ states. One potential explanation for this longer lifetime is that the $4d^{-1}6s$ and $4d^{-1}6d$ states decay by some mechanism other than a spectator transition. To explore this possibility, photoionization calculations were employed to extract the decay channels contributing to the total line widths for all $4d^{-1}n\ell$ states. The result of these calculations support the original hypothesis that all core-excited states with $4d$-character decay via a spectator mechanism on the order of $\sim$6~fs. 

Another possible explanation for the longer lifetime is that the dark-state FWM signatures are actually measuring the lifetime of some other nearby, longer-lived dark state. This idea was explored with a few-level TDSE calculation, in which the inclusion of one additional dark state with an input lifetime of 20 fs yielded simulated FWM spectra in reasonable agreement with the experiment. The TDSE simulations demonstrate that without this additional dark state, the experimental FWM dark-state signatures would have shown $\sim$6~fs decay, indicative of the $4d^{-1}6s$ and $4d^{-1}6d$ spectator decay mechanism. However, due to population cycling with the additional dark state, the FWM signal is extended to as long as the longest lifetime of any populated state. Because the dark-state FWM signature contains a longer lifetime, this additional dark state is most likely one with non-$4d$ character, for example any multi-electronic excited dark state within the 60 -- 70 eV energy region that can be coupled with the NIR light. \textit{Ab-initio} calculations suggest that multi-electron excited dark states with non-$4d$ character are present in this energy region and are potential candidates for such additional longer-lived dark states.

This study constitutes a motivation for theory to identify respective electronic states of non-$4d$-hole character and calculate their lifetimes employing relativistic atomic codes. 
Such insights into the complex electronic structure of highly-excited xenon are important to further our understanding of relativistic many-electron systems, and might lead to new ultrafast control schemes for quantum dynamics in spin-orbit-split systems.\cite{rupprecht2022laser}
Furthermore, characterizing few-femtosecond core-excited state lifetimes directly in the time domain leads the way to attosecond FWM experiments measuring the quantum dynamics of core-level excited states in molecular systems.\cite{rupprecht2025tracing} Exemplified here, attosecond FWM spectroscopy can directly probe core-excited state decay dynamics of complex, many-electronic systems such as xenon on the shortest timescales and with quantum-state-specificity, and, when paired with theoretical support to identify the underlying mechanisms, provides new physical insight into ultrafast core-level decay dynamics.

\section{Supplementary Material}
\label{sec:supplementary_material}

See the supplementary material for a literature summary of the natural linewidths reported for the $4d_{5/2}^{9}5s^{2}5p^{6}[np]$ and $4d_{3/2}^{9}5s^{2}5p^{6}[np]$ Rydberg series, pulse characterization for the two NIR pulses in the FWM experiment, additional FWM time delay scans of the $4d^{-1}_{3/2}6p$, $4d^{-1}_{3/2}6d$, $4d^{-1}_{5/2}6s$, and $4d^{-1}_{3/2}6s$ states, detailed overview of the line widths obtained from the photoionization calculations, and TDSE simulation parameters.

\section{Author Declaration}
\label{sec:author_declaration}

\subsection*{Conflict of Interest}
The authors have no conflicts to disclose.

\subsection*{Author Contributions}
\textbf{Nicolette G. Puskar}: Conceptualization (lead); Data curation (lead); Formal analysis (lead); Investigation (lead); Methodology (lead); Validation (lead); Visualization (lead); Writing– original draft (lead); Writing– review \& editing (lead).
\textbf{Patrick Rupprecht}: Conceptualization (lead); Data curation (lead); Formal analysis (lead); Investigation (lead); Methodology (lead); Software (lead); Validation (lead); Visualization (lead); Writing– original draft (lead); Writing– review \& editing (equal).
\textbf{Jan Dvo\v{r}\'{a}k}: Data curation (lead); Formal analysis (lead); Investigation (lead); Methodology (lead); Validation (lead); Visualization (lead); Writing– original draft (lead); Writing– review \& editing (equal).
\textbf{Yen-Cheng Lin}: Conceptualization (equal); Investigation (equal); Methodology (equal).
\textbf{Avery E. Greene}: Data curation (supporting); Investigation (supporting); Validation (supporting). 
\textbf{Robert R. Lucchese}: Funding Acquisition (equal); Methodology (equal); Project Administration (equal); Resources (equal); Supervision (equal); Writing– review \& editing (equal).
\textbf{C. William McCurdy}: Funding Acquisition (equal); Methodology (equal); Project Administration (equal); Resources (equal); Supervision (equal); Writing– review \& editing (equal).
\textbf{Stephen R. Leone}: Conceptualization (equal); Funding Acquisition (equal); Methodology (equal); Project Administration (equal); Resources (equal); Supervision (equal); Writing– review \& editing (equal).
\textbf{Daniel M. Neumark}: Conceptualization (equal); Funding Acquisition (equal); Methodology (equal); Project Administration (equal); Resources (equal); Supervision (equal); Writing– review \& editing (equal).

\section{Data Availability Statement}
\label{sec:data_availability_statement}
The data that supports the findings of this study are available from the corresponding author upon reasonable request.

\begin{acknowledgments}
We thank C. Leon M. Petersson, Eva Lindroth, and Luca Argenti for fruitful discussions. 

This work was performed by personnel and equipment supported by the Office of Science, Office of Basic Energy Sciences through the Atomic, Molecular, and Optical Sciences Program of the Division of Chemical Sciences, Geosciences, and Biosciences of the U.S. Department of Energy (DOE) at Lawrence Berkeley National Laboratory under Contract No. DE-AC02-05CH11231. N.G.P. acknowledges additional funding from Soroptimist International of the Americas (Founder Region Fellowship). P.R. acknowledges additional funding by the Alexander von Humboldt Foundation (Feodor Lynen Fellowship).
The calculations were mostly performed using the Lawrencium computational cluster resource provided by the IT Division at the
Lawrence Berkeley National Laboratory (supported by the same DOE contract as above).
\end{acknowledgments}


%
%

%



\bibliography{xenon_bib}

\end{document}